\documentclass[pre,twocolumn]{revtex4}
\usepackage[tbtags]{amsmath}
\usepackage{amsfonts}
\usepackage{amssymb}
\usepackage{graphicx}
\renewcommand{\ni}{\noindent}
\newcommand{\tx}[1]{\mathrm{#1}}
\begin{document}
\input psfig.sty

\title{Tradeoff between short-term and long-term adaptation in a changing environment}
\author{Robert Forster$^1$}
\author{Claus O. Wilke$^{2,1}$}
 \affiliation{(1) Digital Life Laboratory, California Institute of Technology, Pasadena, CA 91125}
 \affiliation{(2) Keck Graduate Institute of Applied Life Sciences, 535 Watson Drive, Claremont, CA 91711}
\date{\today}

\begin{abstract}

We investigate the competition dynamics of two microbial or viral strains that live in an environment that switches periodically
between two states. One of the strains is adapted to the long-term environment, but pays a short-term cost, while the other is
adapted to the short-term environment and pays a cost in the long term. We explore the tradeoff between these alternative
strategies in extensive numerical simulations, and present a simple analytic model that can predict the outcome of these
competitions as a function of the mutation rate and the time scale of the environmental changes. Our model is relevant for
arboviruses, which alternate between different host species on a regular basis.

\end{abstract}

\maketitle

\section{introduction}

The quasispecies model \cite{EigenSchuster79} is the premier model to study the evolution of asexual replicators, such as self-replicating molecules or viruses \cite{Eigenetal88,Domingoetal2001}. Originally formulated for constant environments, the quasispecies model has recently been extended to describe adaptation to a changing environment \cite{NilssonSnoad2000, Wilkeetal2001a, NilssonSnoad2002b, KampBornholdt2002, BrumerShakhnovich2004}. The various extensions of the model all work within the original deterministic framework developed by Eigen and coworkers, and thus assume that the population size is infinite. This assumption implies that a population can never lose any genetic information. However, the loss of genetic material, and the mechanisms that prevent it from occurring, are probably major forces shaping the evolutionary dynamics of finite populations in time-dependent environments: A finite population can lose to mutation pressure previously useful genetic material that has become meaningless after a change in the environment, and the population may not be able to reacquire this material when the environment changes back to its original state. Consequently, the population will be at a selective disadvantage in comparison to another population that has managed to prevent a similar loss, even if the second population had to pay some short-term cost to keep the useless genetic material. Therefore, in an environment that alternates between two (or more) states, natural selection faces two conflicting agendas---specialization to the current state of the environment, or adaptation to the long term environment which includes both environmental states.

Here, we study the evolutionary dynamics of competing finite populations of asexual replicators in an environment that alternates between two states, remaining in each state for a time interval of length $T/2$ before switching to the other. To study the tradeoff between the competing forces of selection in this environment, we consider two different strains of replicators, as previously proposed \cite{Wilkeetal2005}. The first strain, which we refer to as the \emph{fused strain}, has a single gene that performs equally well in both environmental states. The second strain, which we refer to as the \emph{divided strain}, has two genes, each of which is advantageous in one environmental state and useless in the other. Clearly, if the fused strain performs as well as the divided strain in both environments, without paying any additional cost, then the divided strain cannot have a selective advantage over the fused strain, regardless of the time scale on which environmental changes happen. If, however, the fused strain does pay some small cost, then which strain is advantageous depends on the exact interplay of the cost, the time scale of environmental change, and the mutation rate. Here, we develop a method to assess and analyze this interplay, and to predict which strain is advantageous in a given setting. As cost, we consider the differential mutation pressure that arises when the fused and the divided strain have genes of different length \cite{Wilkeetal2001b,WilkeAdami2003}. However, it is straightforward to extend our approach to other types of costs.

Note that while we refer to separate genes throughout this paper, our model can also apply to separate functions carried out by a
single gene. In this case, the divided strain corresponds to a gene that can adapt to either function, but not to both at the
same time, whereas the fused strain corresponds to a gene that can adapt to both functions at the same time. Such a situation has
been observed in an artificial-life simulation with a changing environment \cite{LiWilke2004}, where the fusion of genetic
function evolved presumably through changes in the amount of epistatic interactions among the different parts of the organisms'
genomes.

\section{MATERIALS AND METHODS}
\subsection{Model}

Here, we model the evolutionary dynamics of a finite population in a time-dependent environment. For comparison, for an \emph{inifite population} in a time-dependent environment, the quasispecies equation reads \cite{Wilkeetal2001a}
\begin{equation}\label{eq:quasispecies}
 \frac{dy_i(t)}{dt} = \sum_j w_j(t) \mu_{ij} y_j(t) - y_i(t)\sum_j w_j(t) y_j(t)\,, 
\end{equation}
where $y_i$ is the fraction of type $i$ in the population, $w_i(t)$ is the replication rate (i.e., fitness) of type $i$ at time $t$, and $\mu_{ij}$ is the mutation rate per unit time from type $j$ to type $i$. The quadratic term corresponds to the total production of new organisms per unit time, and is subtracted to keep the $y_i(t)$ normalized.

We represent all genes as binary strings. The divided strain has two genes, each of length $L_{\tx{div}}$, and the fused strain has a single gene of length $L_{\tx{fuse}}$. For each gene, there exists a single functional sequence (the master sequence) that confers the selective advantage, and all alternative sequences are non-functional, regardless of the environment.  The reproductive fitness $w_i(t)$ of an individual is determined by whether the individual has a functional gene specialized for the current state of the environment. An individual with the correct functional gene has fitness $1+s$, while an individual without such gene has fitness 1.  Mutations occur upon reproduction with a per-site probability $\mu$, corresponding to a per-gene mutation rate of $U=\mu L$.

\subsection{\label{sec:sim}Simulation}

Both the speed of environmental change and the mutation rate $\mu$ are important factors in determining the outcome of the
competition between the divided and fused strains.  To assess their relative importance, we simulated a population of $N=1000$
individuals reproducing in discrete generations, and with probability of reproduction proportional to their fitness
(Wright-Fisher sampling).  Initially the population was divided equally between fully functional members of the two strains and
the simulation continued until one strain became extinct.  We fixed the length of the divided strain at $L_{\tx{div}}=5$, while
$L_{\tx{fuse}}$ varied from 3 to 11.  We performed $10,000$ replicates at each pair of period lengths and mutation rates, for
$T/2 \ (10,30,100,300,1000,3000,10000)$ and $U_{\tx{div}}=\mu L_{\tx{div}} \ (0.0001, 0.0003, 0.001, 0.003, 0.01, 0.03, 0.1, 0.3,
1, 3)$.

\section{RESULTS}

\subsection{\label{sec:time}Time Scales}

To understand the dynamics of competition between the divided and fused strains in a finite population, we consider two time scales---the competition time scale $T_{\tx c}$ and the drift time scale $T_{\tx{dr}}$. In order to calculate these time scales, we need to know the selective advantage of one strain over the other. The selective advantage $s$ of genotype $i$ over genotype $j$ is a key quantity in theoretical population genetics, and is defined as $s=(w_i-w_j)/w_j$ \cite{CrowKimura70}. While this definition is \emph{a priori} applicable only to individual genotypes, it turns out that to a good approximation standard results from population genetics can be applied to separate strains (which consist of a mixture of closely related mutants) if we treat each strain as an individual genotype with fitness given by the strain average $\langle w\rangle$ \cite{Wilke2001b}. Throughout this paper, we refer to the selective advantage of one strain over another as the effective selective advantage of this strain. We define the effective fitness advantage of the fused strain over the divided strain by \begin{equation}\label{eq:seff} s_{\tx{eff}} = \frac{\langle
w_{\tx{fuse}}\rangle-\langle w_{\tx{div}}\rangle}{\min\{\langle w_{\tx{fuse}}\rangle, \langle w_{\tx{div}}\rangle\}}.  
\end{equation}
This definition guarantees that the magnitude of $s_{\tx{eff}}$ corresponds to the fitness advantage of the superior strain, while the sign indicates whether the fused strain is superior (positive $s_{\tx{eff}}$) or inferior (negative $s_{\tx{eff}}$).  If one strain has an effective fitness advantage $|s_{\tx{eff}}|$ over the other, the competition time scale $T_{\tx{c}}$ is defined to be the typical time until extinction of the inferior strain in a constant environment.  Neglecting finite population
effects, and applying Eq.~\eqref{eq:quasispecies} to strains rather than genotypes, we find that the population fraction $x(t)$ of the superior strain changes approximately according to the logistic equation
\begin{equation}\label{eq:xt}
\dot{x}(t) = |s_{\tx{eff}}| x(t) [1-x(t)],
\end{equation}
subject to our initial condition that $x(0)=1/2$.  To determine the typical extinction time of the inferior strain, we solve
Eq~\ref{eq:xt} for the time when a single member remains of the inferior strain.  Thus
\begin{equation} 
T_{\tx c} = \frac{\ln(N-1)}{ |s_{\tx{eff}}|} \approx \frac{\ln N}{ |s_{\tx{eff}}|}.  
\end{equation}

The drift time scale is defined as the average time for a neutral mutation to go to fixation.  Neutral drift becomes important
when the fitness advantage between the competing strains is small compared to the fluctuations due to finite sampling effects,
i.e. when $s_{\tx{eff}}\lesssim 1/N$ \cite{Kimura64}. For our initial conditions, diffusion theory \cite{KimuraOhta69} predicts
that \begin{equation} T_{\tx{dr}} = 2 N \ln 2 \approx 1390 \, \tx{generations}. \end{equation}

\subsection{Quasispecies Effects}

\begin{figure*}
\centerline{\includegraphics[width=6in]{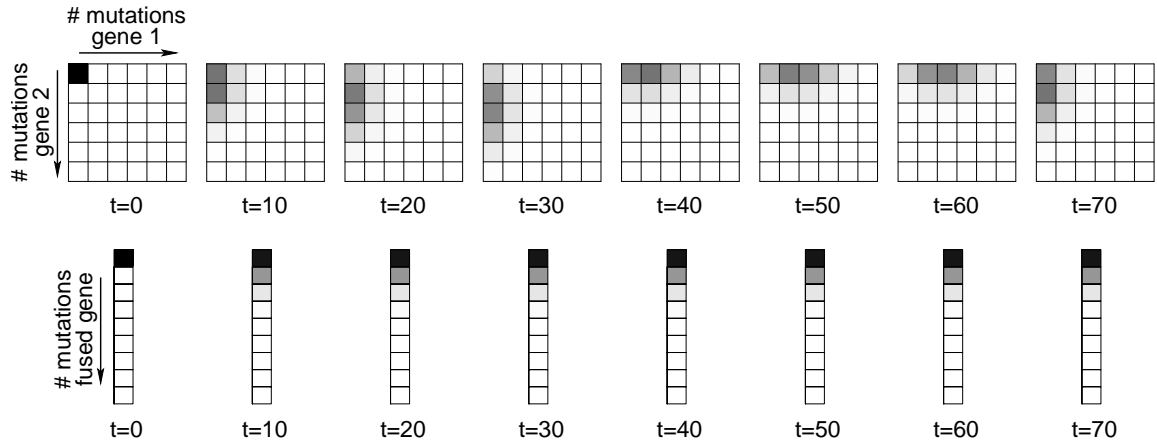}}

\caption{\label{fig:densityplot} Population structure of the divided and fused strains at various time points.  Gray levels
indicate the fraction of sequences at the given mutational distance from the respective error-free sequence. Parameters are:
Oscillation period $T = 60$, per-site mutation rate $\mu = 0.02$, length of a single gene of the divided strain $L_{\tx{div}} =
5$, length of the fused gene $L_{\tx{fused}} = 8$, selective advantage of functional gene $s = 1$, infinite population size.  
From \cite{Wilkeetal2005}.}

\end{figure*}

Given sufficient time to reach equilibrium, each strain will adopt a quasispecies distribution consisting primarily of those members with a functioning gene adapted to the current environment, together with the deleterious mutants that are constantly regenerated through mutation pressure.  In our dynamic fitness landscape, the fused strain will attain this equilibrium distribution after some time, while the divided strain will attempt to equilibrate to the current environment and then go through a period of transition when the environment changes.  As an example, Figure~\ref{fig:densityplot} illustrates the formation of the fused-strain quasispecies and the dynamics of the divided-strain quasispecies in the limit of large population size.

Before we can use the time scales derived in the preceeding section to predict the competition's outcome, we must estimate the average fitness of each strain
appearing in Eq~\ref{eq:seff}.  In our model, the average fitness $\langle w\rangle$ of a strain is given by $\langle
w\rangle=1+s y_0$, where $y_0$ is the fraction of the population that has a functional gene for the current environment.  Thus
Eq~\ref{eq:seff} becomes \begin{equation} \label{eq:seff2} s_{\tx{eff}} = \frac{s y_{0, \tx{fuse}}-s y_{0, \tx{div}}}{1+s
\min\{y_{0, \tx{fuse}}, y_{0,\tx{div}}\}}\,.  \end{equation}

\ni To estimate $y_0$, we assume that the quasispecies immediately reaches its equilibrium distribution.  Denoting $y_i(t)$ as
the population fraction of a given strain with $i$ errors at time $t$, we neglect the back mutation term (that is, the term that represents mutations from genotypes with errors to error-free genotypes) in Eq.~\eqref{eq:quasispecies} and obtain
\begin{equation}\label{eq:quasi} 
\dot y_0(t) = (1+s)Q_{00}(L) y_0(t) - [1+s y_0(t)] y_0(t),
\end{equation}

\ni where $Q_{ij}(L)$ is the probability that a string of length $L$ with $j$ errors mutates into one with $i$ errors. $Q_{ij}(L)$ has been given for example in Ref.~\cite{WoodcockHiggs96}.  Setting $\dot y_0=0$ in equilibrium, we find $y_0 =
[(1+s)Q_{00}(L)-1]/s$.  However, when back mutations become significant and $y_0$ approaches zero in this expression, we reach
the classical error threshold \cite{EigenSchuster79}.  For mutation rates beyond this point, we assume that the population is
randomized uniformly over all possible states, and hence we use
\begin{equation}\label{eq:y0}
 y_0 = \max\Big\{\frac{(1+s)Q_{00}(L)-1}{s}, 2^{-L} \Big\}
\end{equation}

\ni for all mutation rates.  For mutation rates below the error threshold, Eq.~\eqref{eq:y0} yields a simple form for the
magnitude of the fitness advantage, $|s_{\tx{eff}}| = \left(1-\mu\right)^{-|L_{\tx{div}}-L_{\tx{fuse}}|}-1$, while the sign of
$s_{\tx{eff}}$ is given by $\tx{sgn}(L_{\tx{div}}-L_{\tx{fuse}})$.  This result shows that the effect of mutational load on
fitness is to favor whichever strain has the shorter length.

The result of Eq~\eqref{eq:y0} applies to a quasispecies that has reached equilibrium.  While the equilibrium assumption provides a good estimate for the fused strain, the rate of environmental changes may prevent the divided strain's quasispecies from ever reaching this equilibrium.  If the environment changes quickly (relative to the competition and drift time scales), the divided strain persists in an average environment that requires both genes for functionality \cite{Wilkeetal2001a}, and hence this strain has an effective gene length of $2L_{\tx{div}}$.  In this case, we approximate the resulting quasispecies as one with a single gene of length $2 L_{\tx{div}}$ and replace $Q_{00}(L_{\tx{div}})$
by $Q_{00}(2L_{\tx{div}})$ in Eq.~\eqref{eq:y0}.  Even though this approximation disregards the two dimensional nature of the divided-strain quasispecies, it gives a reasonably good estimate of the true $y_0$ for most mutation rates. Note, however, that we do not replace $2^{-L{\tx{div}}}$ by $2^{-2L{\tx{div}}}$.  At any point in time, the environment favors only one of the divided strain's two genes, and hence beyond the error threshold the probability that a randomly-chosen individual from the divided strain carries a functional gene remains $2^{-L{\tx{div}}}$.  We refer to $s_{\tx{eff}}$ calculated with $Q_{00}(L_{\tx{div}})$ as the short-term limit, and to $s_{\tx{eff}}$ calculated with $Q_{00}(2L_{\tx{div}})$ as the long-term limit.

\subsection{Predicting the probability of fixation}

We propose a simple ternary model to predict the probability of fixation $p$ of the fused strain.  In our model, $p$ is $0$ if
the divided strain is favored, $1/2$ for neutral evolution, when both strains are equally likely to go to fixation, or $1$ if the
fused strain is favored.  First, we classify the selective regime based on the drift time $T_{\tx{dr}}$ and the short term
competitive time scale $T_{\tx c} = \ln N / |s_{\tx{eff}}^{\tx{short}}|$ in comparison to $T/2$, the length of time for which the
environment remains constant (see Table~\ref{tab:sel-reg}). If $T_{\tx c}<T/2$, then we expect the competition between the two
strains to end before the environment changes even once, and hence the short-term limit applies. The value of $T_{\tx{dr}}$ is
irrelevant in this case.  If both times are longer than $T/2$, then we expect the competition to extend over several
half-periods, and hence the long-term limit applies. Finally, if $T/2$ is smaller than $T_{\tx c}$, but larger than
$T_{\tx{dr}}$, then we expect drift to be the dominant force. We call this regime the neutral limit, and set $s_{\tx{eff}}=0$.
Having determined the appropriate limiting case (short-term, neutral, or long-term limit), we can use the associated fitness
advantage $s_{\tx{eff}}$ to predict the probability $p$ of fixation for the fused strain (Table~\ref{tab:model-pred}). If
$|s_{\tx{eff}}| < 1/N$, then the two strains are effectively neutral. Hence, the outcome of the competition is determined by
drift, and $p=1/2$. Otherwise, $p=0$ or 1 depending on whether $s_{\tx{eff}}$ is negative or positive.

\subsection{Comparison with simulation results}

From our simulations, we estimated the probability $p$ that the fused strain would fixate in the population as a function of $T/2$, $U_{\tx{div}}$, and $L_{\tx{fuse}}$.  Results are shown in Figure~\ref{fig:results}A, B, C, for representative values of $L_{\tx{fuse}}$.  The probabilities obtained by simulation have a standard error of $\pm 1\%$. Figure~\ref{fig:results}D, E, F shows the corresponding predictions of our model.  Over the full range of parameters described in Section~\ref{sec:sim}, our model predicted that the divided strain was superior in 235 cases ($p=0$), that the strains were neutral in 240 cases ($p=1/2$), and that the fused strain was superior in 155 cases ($p=1$).  The range of simulation results corresponding to each of these three predictions is shown in Figure~\ref{fig:error}.  When the model predicted $p=0$, $85\%$ of all simulation probabilities fell in the range $0-0.1$.  When $p=1$ was predicted, $84\%$ of simulation probabilities fell between $0.9-1$, while in the neutral case ($p=1/2$ predicted), $58\%$ of simulation probabilities fell between $0.45-0.55$.  The rms error between the simulation values and the model predictions is $18.6\%$, averaged over all cases.  For comparison, the best possible rms error for any such ternary model is $7.6\%$ on this data.

\section{DISCUSSION}

The study of quasispecies dynamics in a time dependent fitness landscape to date has primarily focused on the limit of infinite population size \cite{NilssonSnoad2000, Wilkeetal2001a, NilssonSnoad2002b, KampBornholdt2002, BrumerShakhnovich2004}.  In a periodic fitness landscape, an infinite population size guarantees that competition between two strains will result in the deterministic extinction of the inferior strain or, in certain finely tuned cases, an unstable coexistence between the strains (although frequency dependent selection may stabilize this equilibrium \cite{Wilkeetal2004}).  In contrast, the generalization of these models to a finite population presents a continuous range of possibilities from almost certain extinction
to the complete randomness of neutral drift.

\begin{figure*}
\centerline{\includegraphics[width=6in]{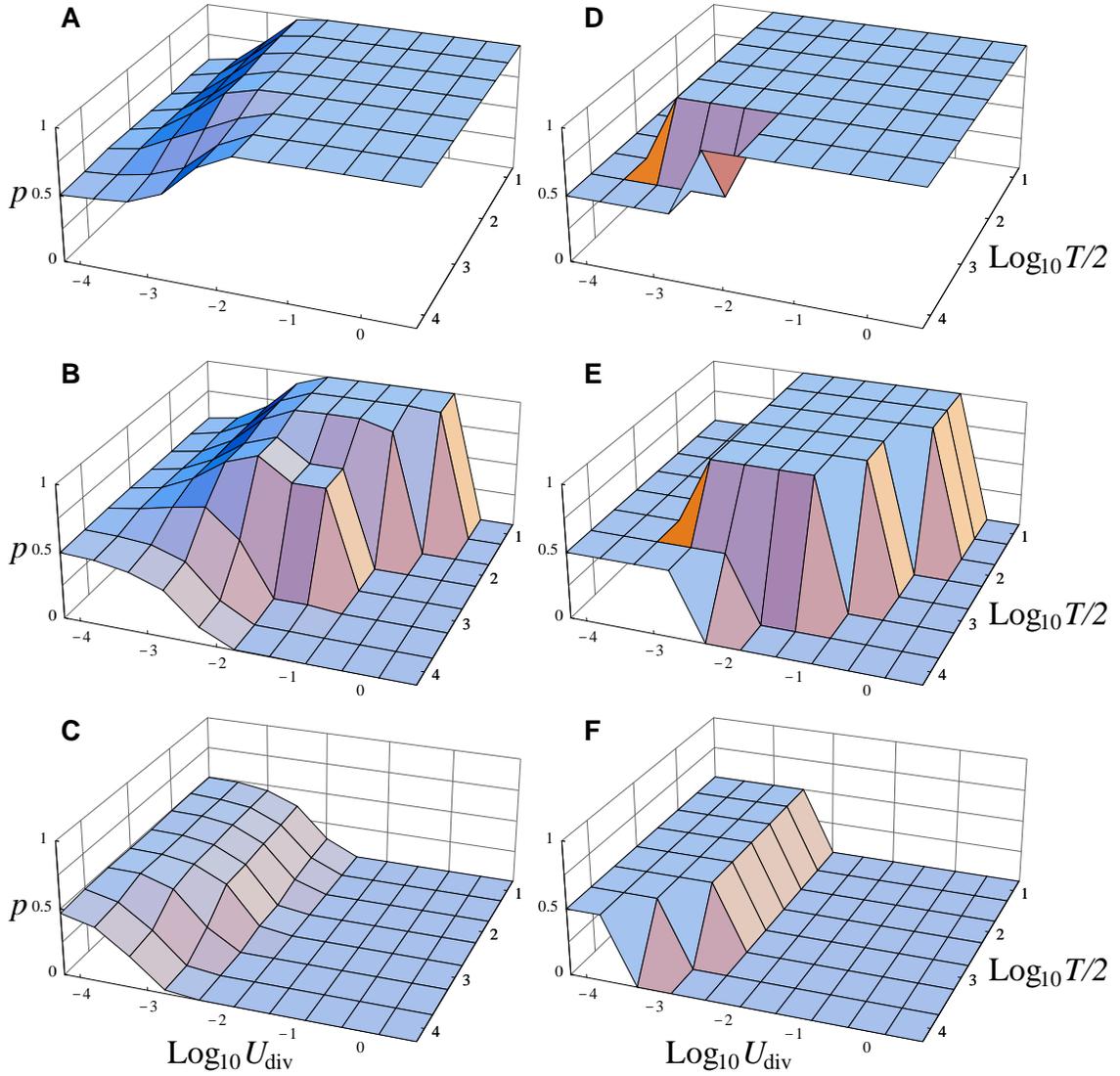}}

\caption{\label{fig:results} (Color online) Left (A, B, C): Simulation results for the probability of fixation of the fused strain as a function of the mutation rate $U_{\tx{div}}=\mu L_{\tx{div}}$ and period length $T$ for $L_{\tx{div}}=5$ and $L_{\tx{fuse}}=4,6,11$ (top to bottom).  Simulation results have a standard error of approximately $\pm 1\%$.  Right (D, E, F): Model predictions for the
same.  Parameter values are $s=1$, $N=1000$.}

\end{figure*}

\begin{figure*}
\centerline{\includegraphics[width=6in]{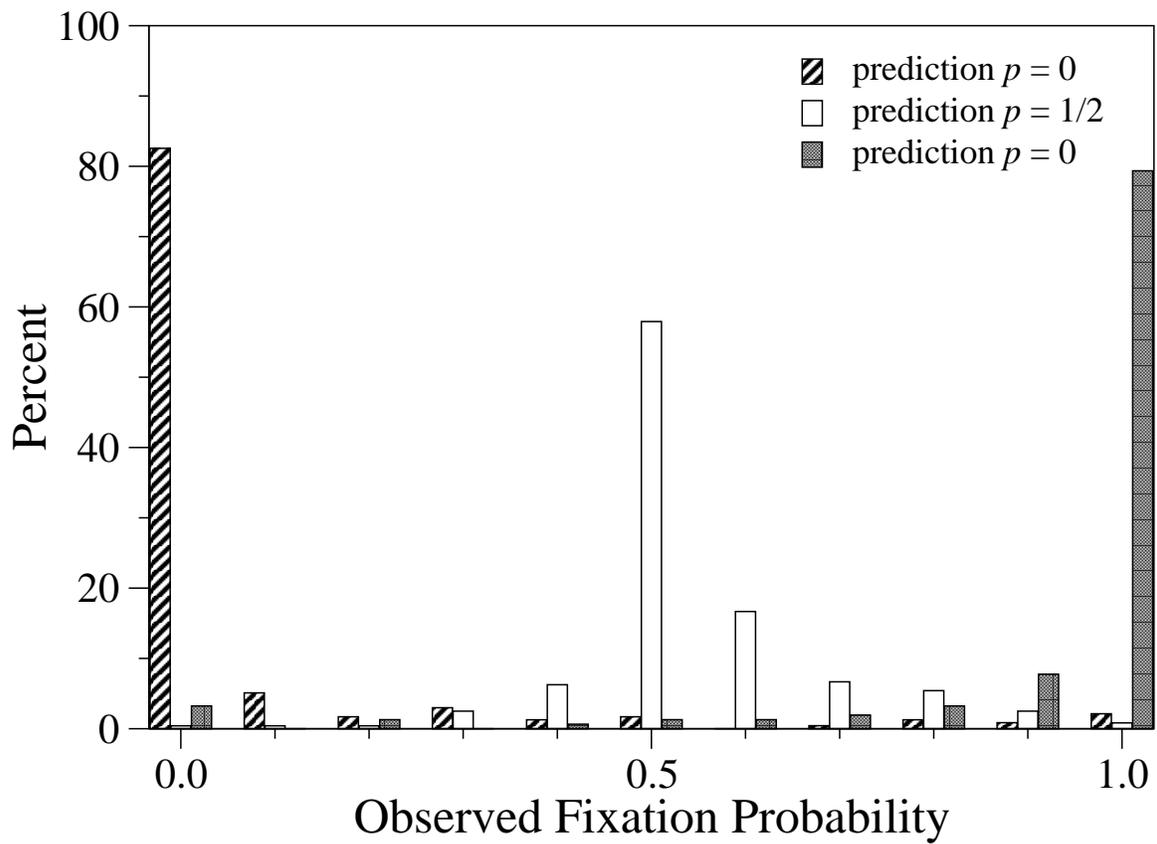}}

\caption{\label{fig:error} Fixation probabilities for the fused strain (as determined by simulation), classified by model prediction.  The model predicted $p=0$ in 235 cases, $p=1/2$ in 240 cases, and $p=1$ in 155 cases.  The $x$ axis is binned in $0.1$ increments, with the bins' midpoint value shown.}

\end{figure*}

In this study, we present a model for predicting the outcome of competition between finite quasispecies' in a periodic
environment.  As applied to our specific case of competition between a divided and a fused strain, our model shows good
qualitative and quantitative results in comparison with simulation (Figs.~\ref{fig:results} and~\ref{fig:error}).  When
$L_{\tx{fuse}}<L_{\tx{div}}$ or $L_{\tx{fuse}}>2L_{\tx{div}}$, one of the two strains is strictly superior, and the outcome of
the competition is determined by whether drift or competition are more important at the given mutation rate.  For intermediate
values of $L_{\tx{fuse}}$ the competitive dynamics are more complex, as certain combinations of mutation rate and period lengths
favor the fused strain while others favor the divided strain.

The only qualitative feature of the competitive dynamics we have ignored is the time for the quasispecies to reach its
equilibrium distribution.  We decided not to include a quasispecies time scale in our model after preliminary efforts showed that
the added complexity failed to significantly improve model accuracy.  While the fused strain quickly reaches its equilibrium
independent of the environment's state, the divided strain may fare worse than predicted in some special cases when it spends
significant time in transitions between the quasispecies distributions for each environment, being poorly adapted in the
meanwhile.  Still, these cases are relatively rare since at low mutation rates quasispecies effects are less important, while at
high mutation rates the error threshold masks any such effects.

Our results suggest that, under appropriate circumstances, a selective pressure exists to fuse or divide complementary genes in a periodic environment.  The tendency of uncertain enviromental conditions to facilitate large scale genetic changes such as this one has recently been studied \cite{EarlDeem2004}.  Most genomes are full of apparently useless or non-functional genetic material, which in
our model corresponds to the excess length of the fused strain over a single gene of the divided strain.  In populations as small
as 1000 such as those we studied, such temporarily useless genetic material comprising $20\%$ of the genome could be stably
maintained for periods of dormancy up to 1000 generations (Figure~\ref{fig:results}B).  Alternatively, a fused gene exposed to an
appropriate periodic environment might undergo division in response to the selective pressure we describe.  The beginnings of a
genome segmentation has been observed in foot-and-mouth disease virus (FMDV) in response to conditions of high multiplicity of
infection \cite{Garciaetal2004}.  One explanation of this segmentation is that translational speed favors shorter genes
\cite{Nee1987, Szathmary1992}, a fitness effect similar to the length-based mutational pressure of our model.

Our model applies directly to the evolution of arboviruses, which are viruses transmitted by arthropods. For example, the arbovirus West-Nile virus is transmitted from birds to birds (and the occasional human) by mosquitoes, experiencing the alternating environments of the avian and insect hosts. Experimental virologists have long tried to determine whether viruses subjected to such alternating environments adapt to the short-term or the long-term environment, but have not found a conclusive answer.
Experiments with vesicular stomatitis virus (VSV), eastern equine encephalitis virus (EEEV), and Dengue virus in cells of insect
and mammalian origin have shown that in some cases, adaptation to one cell type leads to loss of fitness in the other cell type,
while in other cases fitness can increase in both cell types at the same time \cite{Novellaetal95, Novellaetal99, Weaveretal99,
TurnerElena2000, CooperScott2001, Chenetal2003, ZarateNovella2004}. Which of the two cases occurs depends on the time spent in
each of the two hosts, and also on the details of the fitness landscape in the two hosts. It stands to reason that in future
experiments in which the time scale of environmental change is varied over a wide range, a switch in the adaptation strategy from
short term to long term will be observed, and that the time scale at which the switch occurs can be predicted with the methods we
have developed here.

In conclusion, we have demonstrated that the competitive dynamics of finite populations in a time-dependent environment can be
quite complex, but that nevertheless, estimates of the effective fitness advantages of the different strains together with an
understanding of the drift and competitive time scales can lead to remarkably accurate predictions of the evolutionary dynamics.
We believe that similar techniques will prove useful to interpret and predict outcomes of virus-evolution experiments in changing
environments.


\cleardoublepage

\begin{table}
\caption{\label{tab:sel-reg}Selective regime, as determined by the relative magnitudes of $T_{\tx{dr}}$, $T_{\tx c}$, and $T/2$. }
\begin{center}
\begin{tabular} {c|c} 
Condition & Selective Regime \\ \hline
$T/2>T_{\tx c}$ & short term limit\\ 
$T_{\tx{dr}}<T/2<T_{\tx c}$ & neutral limit\\ 
$T/2<T_{\tx{dr}}\,,T_{\tx c}$ & long term limit\\ 
\end{tabular}
\end{center}
\end{table}

\begin{table}
\caption{\label{tab:model-pred}Model predictions, as determined by the relative magnitude of $s_{\tx{eff}}$ and $1/N$.}
\begin{center}
\begin{tabular} {c|c|c} 
Fitness Advantage (Fused Strain) & Prediction & $p$\\ \hline 
$s_{\tx{eff}}\le-1/N$ & divided strain wins & 0 \\
$-1/N<s_{\tx{eff}}<1/N$ & neutral$^\ast$ & 1/2 \\
$s_{\tx{eff}}\ge1/N$ & fused strain wins & 1 \\ 
\end{tabular}
\end{center}
$^\ast$ In the neutral case, both strains are equally likely to win.
\end{table}

\end{document}